A convolutional neural network technique for online tracking of the radius evolution of levitating evaporating microdroplets of pure liquids, liquid mixtures and suspensions

Kwasi Nyandey[a, b, *], Gennadiy Derkachov[a] and Daniel Jakubczyk[a]

[a] Institute of Physics, Polish Academy of Sciences, Al. Lotników 32/46, 02-668 Warsaw, Poland
[b] Laser and Fibre Optics Centre, Department of Physics, School of Physical Sciences, College of Agriculture and Natural Sciences, University of Cape Coast, Cape Coast, Ghana

*Corresponding author: kwasi.nyandey@ucc.edu.gh/nyandey@ifpan.edu.pl

**Abstract**

We have used convolutional neural network in a classification task to track the radius evolution of levitating evaporating microdroplets of pure diethylene glycol, diethylene glycol-water-polystyrene microparticles suspension, dipropylene glycol-water mixture and dipropylene glycol-water-silica nanoparticles suspension. We discretized a wider radii range into short radii segments, labeled them with class numbers and generated theoretically light scattering patterns from Mie theory. Then the network was trained on the theoretical images and used to classify unlabeled experimentally recorded Mie scattering patterns from the evaporating microdroplets. A plot of the class-average-radii versus the camera's time step revealed the profile of the entire droplet radius evolution. We were able to work with ~1500 classes and showed that the technique has the potential to distinguish droplet size difference of ±5 nm. We expect it to be applicable for online/real-time tracking of droplet evaporation.

## 1. Introduction

Evaporation of microdroplets has been already fairly well studied (see e.g. [1–3] and references therein), droplet size (radius) being, of course, the primary parameter of interest. However, the instantaneous optical study of microdroplets' evaporation still remains a challenge and is sought after for the online investigation of unsteady and transient phenomena in droplets' evaporation. The accuracy, precision and speed of data processing make neural networks a promising tool for application in this field as it is already in other fields – real-time facial [4] and object [5] recognition, self-driving cars [6] and sensor-based controls for robots and industry [7]. In this study, our goal is to use a convolutional neural network (trained on theoretically generated Mie scattering patterns, grouped into classes encompassing consecutive small ranges of radii) to classify corresponding experimentally recorded sequences of patterns, enabling the tracking of the radius evolution of evaporating microdroplets of pure liquids, liquid-liquid mixtures, and suspensions.

Numerous sizing methods have been proposed, (including application of neural networks [8]), which naturally belong to two major groups: off-line and online. Off-line methods can be quite accurate – in our laboratory we have previously combined electrodynamic trapping and several optical methods [9–12] for sizing of evaporating microdroplets of pure liquids and liquid suspensions of nanoparticles. By comparing experimental light scattering patterns to a library of corresponding patterns generated from Mie theory [9] – in favorable cases – we were able to determine the droplet radius with the accuracy of ±10 nm for pure droplets. Several, more or less sim-ilar techniques for droplet/particle sizing and tracking were also reported by other authors [1–3,13,14]. Somewhat different techniques, but also making use of interferometric effects exist as well: e.g. a combination of Mie-theory and aerodynamics [15], interferometric laser imaging for droplet/bubble sizing (ILIDS) [16], etc. Wu et al. [17], used phase interferometric particle imaging to measure the size of large fuel droplets (micrometer size range) and their changes (in nanometers) during evaporation. Their experimental measurements were checked versus the theoretical simulations. Some techniques permit measurements in the forward (or near forward) direction (see e.g. [18]), where the scattering pattern is fairly simple and the diameter of the particle can be easily related to the number of fringes or their spacing [16]. Others choose the direction of the rainbow angle [19], where the scattering pattern is more sensitive to the refractive index than to the droplet size. It is worth noting that the interference phenomenon in the illuminated droplet (sphere) can be mathematically described or handled in various ways. For instance, it can be described as the interference produced by light reflected from the surface and refracted through the interior of a (often large) spherical transparent particle [16]. Others have exploited just the interference resulting in the two glare points [16,18]. Whenever the size of the particle falls outside the so-called resonant size-range, geometrical optics approximation can also be applied.

All these methods provide frameworks for accurate droplet sizing, albeit the acceptable computational time required for online sizing, through solving the inverse scattering problem, depends on the rate of droplet evaporation, which restricts them to slowly evaporating droplets. For online/field sizing, various authors have reported several techniques including Fast-Fourier-Transforming the Mie scattering patterns for *in situ* sizing (with 2% experimental uncertainty) [20] and Laser Doppler Velocimetry (LDV), which are essentially independent of Mie-theory. The LDV methods are generally based on dual-scatter Laser Doppler velocimetry and geometrical/physical optics considerations [21–24] for the simultaneous determination of particle size, number density and velocities of typically spherical droplets/particles in fluid flows. A related technique (particle tracking velocimetry) has found application in droplet tracking. The technique involves imaging of particles and using computer algorithms for tracking their trajectories in successive frames [16]. Others have combined the technique with ILIDS for tracking and sizing evaporating fuel droplet [25]. The well-known shadowgraphy is another simple technique which has been used for online droplet sizing [26]. These methods [20–24,26] (reported to be easier to use) are fast but less accurate, so there seems to be always a trade-off between fast processing and obtaining high accuracies.

While diverse techniques can be used for droplet radius measurements, a complete study of droplet evaporation requires thermodynamic models (see e.g. [27] and perhaps [28] for comparison), which rely on the detailed analysis of temperature, density and vapor pressure profiles, as we discussed in earlier publications (e.g. [29,30]). Since droplet radius can also be obtained from these thermodynamic models, radius evolution curves from measurements are often compared with a chosen thermodynamic model, which appropriately represents the experimental conditions. In this work, we employ the radius-square-law:

$$R^2(t) - R^2(t=0) = At, \qquad (1)$$

as a ground-truth to compare our sizing technique to, where $R$ is the time ($t$) varying radius of the evaporating droplet and $A$ is a constant, which can be sought also with theoretical models.

In this paper, we present a first attempt at developing a technique for characterizing the size of evaporating microdroplets, thereby yielding the temporal evolution of the radius – potentially both fast and with an accuracy as high as few nanometers. We generated theoretical light scattering pattern from Mie-theory to train a convolutional neural network (CNN) and used the trained network to classify: (i) new set of similar light scattering patterns where the radius was generated from the radius-square-law, (ii) experimentally recorded Mie scattering patterns from droplets of pure liquids, binary-liquids mixtures and their suspensions (see experiment section). We kept to solving this problem as a classification task as we did for suspensions in cuvettes [31,32]. In this way, we could possibly use one CNN type to recognize both the suspension type and estimate the microdroplet radius. However, it must be noted that such approach is rather non-canonical and may not be optimal, since looking for a continuously changing value, like radius evolution, naturally calls rather for a regression task with CNN (compare e.g. [33]). However, resolving this issue is beyond the intended scope of this paper.

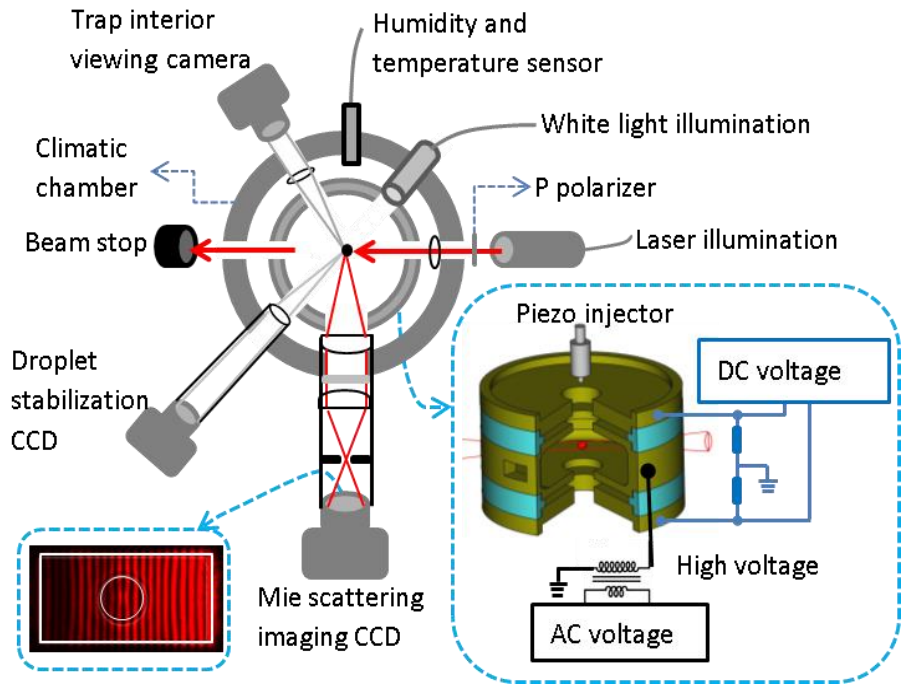

Figure 1: Schematic diagram of the experimental setup (top view). **Right inset:** electrodynamic trap 3D-drawing showing internal structure, wiring, piezo injector, droplet (out-of-scale) and incident wave configuration. **Left inset:** Mie scattering patterns showing a rectangular mask and a circle displayed in the frame processor window.

## 2. Experimentation

An operational method of droplet radius tracking using light scattering must be robust and devoid of experimental uncertainties associated with the beam-droplet-detector misalignment and other geometrical imperfections of the setup. In view of this, a variant of quadrupole electrodynamic trap was developed in our laboratory [34] suitable for angle-resolved elastic light scattering measurements. The details of further modification specific to this work are presented in Figure 1 and described below. The trap was kept in a climatic chamber filled with dry nitrogen at ambient pressure and a constant temperature (±0.2°C change in the course of the experiment), albeit the presented experiments were carried out at several different temperatures. Considering the microdroplet size and the fact that the climatic chamber was pre-stabilized, the temperature of the microdroplet and the climatic chamber could be considered constant. The droplets were introduced on demand into the trap with a piezo injector [35] pre-inserted through the top port (see right inset in Figure 1). Since not all injections result in successful droplet trapping (and recording), before each injection, the climatic chamber was flushed with the dry and filtered nitrogen gas to eliminate remnants of previous injection and other possible contaminants. Still, after several droplets were successfully injected and recorded (across multiple experimental runs), remnants of liquid and particles accumulate on the walls of the trap, and distort the trapping fields. Then the trap must be thoroughly cleaned. The setup had to be disassembled for this cleaning. The construction of the setup, aided by the use of dedicated alignment tools, allowed for reassembly into its original state without any appreciable change in configuration or setup parameters, ensuring that data could be collected reproducibly.

The droplet was kept at the center of the trap by a stabilization loop (see details in [34]) and a plane-polarized red laser (658 nm, ~10 mW power, ~ 1 mm beam diameter and polarized parallel to the scattering plane) was used to illuminate it. We used a 14-bit color camera (Pike F-032C, AVT, 640×480 pixels and 7.4×7.4 μm pixel size – KAI-0340 with HAD microlens CCD) to record the scattering patterns. A confocal imaging system, made up of a 16.07 mm BFL aspherized achromatic lens (12.5mm diameter, 20mm EFL, Edmund Optics) and a 35 mm focal length plano-convex lens respectively, enable the projection of the scattering pattern onto the CCD (Refer to Figure 1). In order to obtain high precision in size measurements, it is critical that the center angle as well as the angle per pixel is correctly determined [36]. In view of this, the solid angle, subtended by a rectangular aperture of height 4.30 mm (and a width of 8.96 mm), was positioned at the center-most part of the CCD sensor (see left-inset in Figure 1). The field of view was centered at the azimuth angle of 90 ± 0.1° to the illumination direction. A rectangular mask and a circle were displayed in the frame processor window (see left-inset in Figure 1). The rectangle defines the region of interest (ROI) and the circle's center, the position of the lens axis respectively on the CCD. The circle was aligned with the Newton ring resulting from the anti-reflection coating on the aspherized lens illuminated from the back. Since the top and bottom edges of the aperture are well defined, the angle per pixel values was determined using the corresponding edges of rectangular ROI and the geometrical relations in the trap. The left and right edges of the rectangular aperture partially overlapped with the circular objective aperture, so the interference pattern azimuthal range was safely defined by the left-right edges of the (subtending) rectangular ROI. It ensured that a good correspondence between the measured pattern and the computed Mie scattering pattern could be found, in other words – the fit was unambiguous. Furthermore, it al-lowed us to crop out well-defined patterns within the images without edge distortions for use with CNN. Hence, the effective field of view determined from the ROI covered an angular span of ± (15.40 ± 0.05°) in opposite directions and an angle of elevation of 5.00 ± 0.05°.

We recorded videos of each trapped droplet at 75 fps, capturing the light scattering patterns throughout the entire evaporation process (droplet evolution). The red channel (CCD RGB filter, well-fitting to the red laser wavelength) from the recorded movies' 640×480-pixel frames was extracted. The scattering patterns within the rectangular ROI were further cropped out into 221×501 (rows and columns respectively) images before normalizing to a 0 – 1 brightness range.

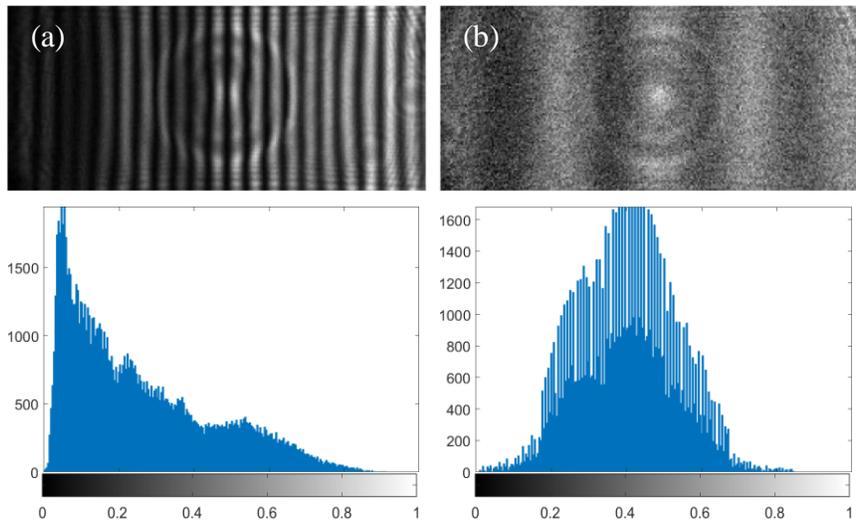

The pure droplets were produced from diethylene glycol (DEG), 99.0% (BioUltra, GC, Fluka). Samples of scattering patterns at the beginning and at the end of the evolution – respectively – of a droplet of pure DEG have been presented in Figure 2 (a) and (b). Their corresponding histograms have been presented below them. The signal-to-noise ratio is higher at the beginning of the evolution, since scattered light intensity increases roughly as the square of the droplet radius. The interference rings seen in the images in Figure 2 are artifacts of the front lens as mentioned above. We extracted radius evolution from the scattering data with our Mie Scattering Lookup Table Method, we described in [9].

Figure 2: Light scattering patterns recorded at the beginning (a) and the end (b) of evolution of a levitated microdroplet of pure diethylene glycol. Their corresponding histograms are presented in the bottom panel. The interference rings seen in the images in the top panel are artifacts of the AR-coated (aspherized) lens illuminated from the back.

The binary liquids mixture was produced by mixing dipropylene glycol (DPG), 99% (mixture of isomers, Alfa Aesar) and distilled water in volume ratio of 6:4. Then, the droplets of suspensions were produced from two suspensions of low concentration: (i) a mixture of DEG and monodispersed aqueous suspension of polystyrene microspheres, in volume ratio of (DEG:PS suspension) ~9:1 and (ii) a mixture of DPG and monodispersed aqueous suspension of silica nanospheres, further diluted with distilled water to a final volume ratio of (DPG:$H_2O$:$SiO_2$) ~5:4:1. Each suspension was prepared by mixing and ultrasonication. The diameter of the PS microspheres (2.5 wt% in manufactured suspension, SIGMA) was 1.1 μm, refractive index ≅ 1.59. The

SiO$_2$ nanospheres (5 wt%, Corpuscular Inc.) had a diameter = 450 nm and refractive index $\cong$ 1.465.

The concentration was specifically chosen so that the Mie scattering from the smooth droplet surface can still be observed but with intermittent interferences (shadow casting/distortions) from the microspheres. These images will be particularly important to our technique, because they allow finding the droplet radius independently with off-line techniques, in which mean scattering patterns are computed from images making them very robust against any partial distortions [9–11]. Observing Mie fringes at the early stage of evolution before the images get completely speckled due to scattering from the microspheres eventually also enables calibration of droplet sizing by electrostatic weighting (compare [9]). On the other hand, our CNN operates by sampling information from the entire images, which also should ascertain our technique's robustness. Verification of this intuition belongs to the main scope of the presented work.

### 2.1. *Theoretical data generation*

Because we intended to train the CNN to both cover a wide range of radii and achieve high accuracy in determining droplet size, we were limited by the largest number of classes the network could support without exceeding our resource capacity. It turned out that 30 µm could still be chosen as the upper limit of droplet radius. On the other hand, we selected the lower droplet radius limit to be 1 µm for compatibility reasons. Of course, in the case of a suspension of PS 1.1-µm-diameter microspheres, this limit will never be reached for reasonable initial number concentrations of microspheres, although it could well be reached for pure droplets (compare [37]). Droplet evaporation monitoring involves the discretization of a time-varying continuous radius anyway, and the goal is usually a possibly smallest discretization step. Hence, instead of allowing a trade-off, we adopted a coarse-to-fine radius classification. Thus, in the coarse radius classification, the radii range (1 – 30 µm) was divided into 967 classes with a class interval of 30 nm. The radii within each class were randomly generated independently using the Matlab random generator (rand) and sorted in descending order. Then theoretical light scattering patterns were generated with the Mie theory-based codes, using the measured angular span (azimuthal), wave properties (wavelength and polarization), a constant refractive index of $1.4455 + 0.0i$ (corresponding to DEG at 658 nm and 21°C) and the varying radius. Each generated pattern was composed into a 221x501 image (corresponding to the angular elevation and azimuthal in the ROI) with brightness values normalized to a $0 - 1$ range (see Figure 4 (a)). Henceforth, all light scattering patterns (generated or recorded) will be referred to as images unless there is a strong need. This normalization ensures that the network is not susceptible to brightness differences between theoretical and experimental images as well as variations within experimental images due to changing droplet size. It also ensures that more emphasis is placed on the width and positions of the peaks in the scattering pattern. We generated 2000 images within each class and divided the whole data into training and validation sets for training the convolutional neural network (see section 3.2 for details). As mentioned earlier, we verified that the generated Mie scattering patterns fit well with the scattering patterns recorded from a microdroplet (see Figure 3).

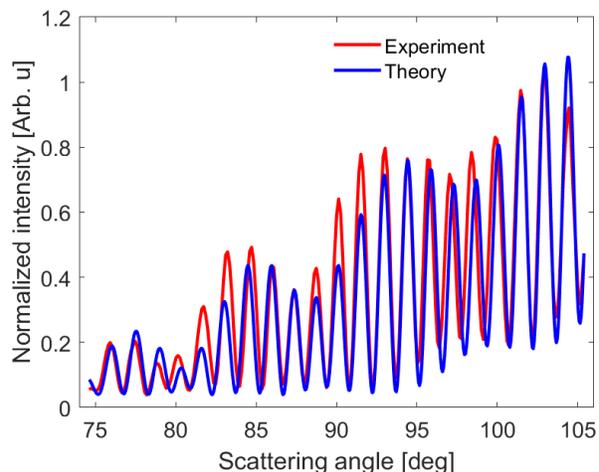

Figure 3: A fit of Mie theory (radius = 12723.3 nm) to the experimental light scattering pattern (compare Figure 2 (a)). The slight discrepancies may be caused by the remnant optical aberrations, finite image acquisition time causing integration over the droplet radius and (unaccounted) non-uniformity of the background, though the deviation of the droplet from sphericity may also contribute.

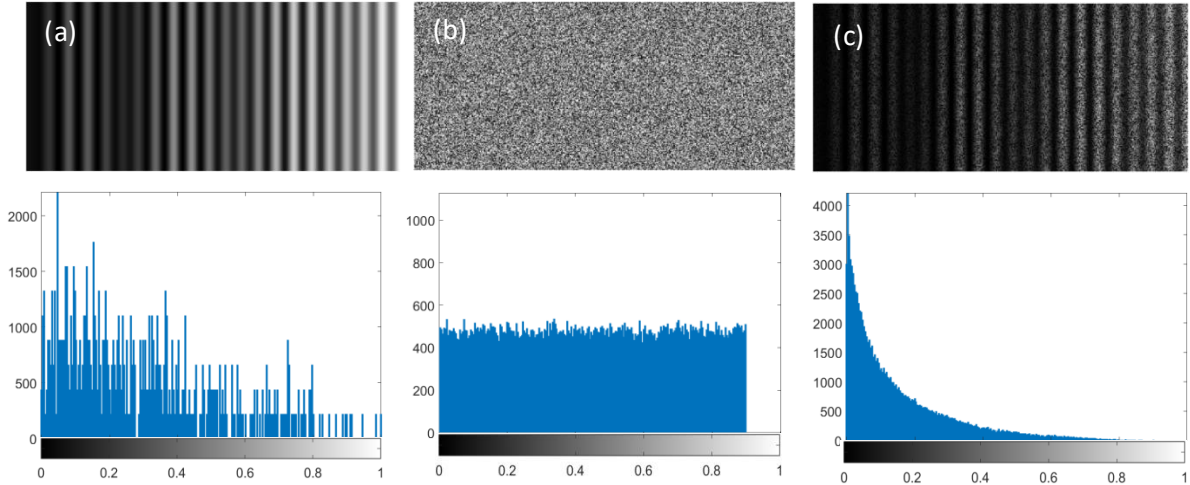

Figure 4: Theoretically generated light scattering pattern (a), uniformly distributed random noise (b) and the result of elementwise multiplication of (a) and (b). Their respective histograms are presented in the lower panel.

The choice of randomly generated radii over a given range is reasonably justified by the Mie theory predictions – patterns (images) that look significantly different from their neighbors in the sequence (movie) correspond to very narrow morphology dependent resonances (MDRs), which appear rather regularly and sparsely versus droplet radius – "long tail" effect associated with such resonances should not, in general, bother us here [38] (we'll comment on that in section 4.4.3). In the above approach, several radii will be classified into one 30-nm class thereby allowing the maximum accuracy of ±15 nm. We achieved an even higher accuracy of ±5 nm (fine-radius classification) by dividing the wider radii range into two sub-regions (1 to 15 µm) and (15 to 30 µm) with a class-interval of 10 nm.

| Data label | Radii range [µm] | Class interval [nm] | Number of images in a class | Number of classes | Network label |
|---|---|---|---|---|---|
| Training data 1 | 1 – 30 | 30 | 2000 | 967 | Stage 1 |
| Training data 2 | 30 – 15 | 10 | 1000 | 1501 | Stage $2_1$ |
| Training data 3 | 15 – 1 | 10 | 1000 | 1401 | Stage $2_2$ |
| Test set | 13 – 1 | – | – | – | – |

Table 1: Data table showing the composition of generated data and labels of trained network.

Henceforth we label the coarse-radius data as *"Training data 1"* with 967 classes and a class interval of 30 nm. For convenience and in relation to evaporating droplet, we reverse the order and label the two sub-regions as *"Training data 2"* (30 – 15 µm) and *"Training data 3"* (15 – 1 µm) with 1501 and 1401 classes – respectively – and a class-interval of 10 nm. For these data sets we generated 1000 images in each class. Then the same procedure, as described above, was followed to generate the datasets in both cases. As mentioned above, the data was generated and processed in Matlab (2023a) environment on a PC equipped with two GPUs (NVIDIA GeForce RTX 3090). However, only one GPU was used to generate the data (images).

Finally, we used equation (1) to generate additional dataset by choosing an initial radius of 13 µm, and an evolution time of 3000 seconds (with a time step of 1 s). We labelled this data as the *"Test set"*. A summary of all the data generated has been presented in section 3.2 with the corresponding labels of the trained network. Considering the fact that the initial minute amount of impurity in a microdroplet and the contribution of electronic noise from the CCD become significantly high when the microdroplet size reduces, we tried to make the theoretical images mimic the experimental images as closely as possible. Thus, we generated 22×501 uniformly distributed random noise patterns (with double-precision floating-point values within the range 0 – 0.9) and multiplied (elementwise) each generated image by the noise independently and normalized to 0 – 1 range. An example of theoretically generated image is shown in Figure 4 (a) and the sample noise in Figure 4 (b). Figure 4 (c) represents a theoretical image which has been multiplied by random noise (compare histogram of Figure 2 (a) and Figure 4 (c)). It is worth noting that we experimented with different noise levels (0, 0.1, 0.3, …, 1.0) and found that the network trained most effectively at a noise level of 0.9, resulting in higher and more stable recognition of the experimental images.

### 3. Network architecture and training.
#### 3.1. *Architecture details*
It is widely known nowadays that the simplest way to enhance the performance of deep neural networks is to increase their depth – number of levels – and their width – number of units. However, this is often associated with the two major problems – overfitting and increased use of computational resource [39]. Since, as mentioned above, we dedicated to the task a PC equipped with two fairly capable GPUs, we were able to increase both the depth and width of the CNN we used in our previous studies (see [31,32,40]) and train the network to recognize over 1500 classes. We increased the number of convolutional layers from 8 to 11, consisting of 9 convolutional layers and 2 skip-convolutional layers (see Figure 5). The first 5 convolutional (*conv1* to *conv5*) and the skip-convolutional layers were unchanged (see description in [31]). We renamed *conv6* as *conv6A* and created a parallel branch named *conv6B*, where both are fed with the learned features from *conv5*. The two branches were extended by connecting *conv7A* and *B*, respectively, to *conv6A* and *B*. A 3×3 filter size with a stride of [1 1] was set in both *conv6A* and *B* while the number of filters were 128 and 256 respectively. In *conv7A* and *conv7B* we used 3×3 filters with stride of [2 2]: 256 and 382 filters respectively. The outputs from both branches were concatenated along the channel dimension into a total of 638 channels using a *depth concatenation* (*D-cat*) layer. Each of the 9 convolutional layers was followed by a *Batch Normalization (BN)* layer and a *Rectified Linear Unit (Relu)* layer. We alternated *Max Pooling* (*MP*) layer between the convolutions, thus *conv1, conv3, conv5 and conv6A* and *B*. Finally, the *depth concatenation* layer is connected to a *classification* (*Classif*) layer through a *fully connected* (*FC*) layer (with the number of neurons corresponding to the number of classes) and a *softmax* (*S-Max*) layer.

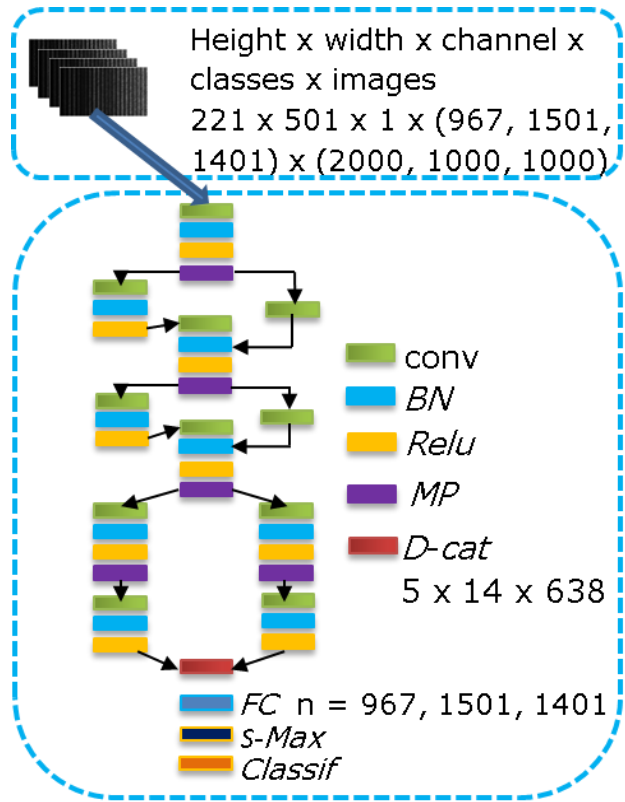

Figure 5: An illustration of our convolutional neural network.

### 3.2. Training methodology

As mentioned earlier, we trained the network on the three independent datasets (see Table 1). Before training, each dataset was split into training set (90%) and validation set (10%) by a randomized selection. The cross-validation was carried out during training to ensure that the network does not fit to noise in the data. We set the cross-validation to be carried out at the end of every epoch and automatic stopping condition for training, when the validation loss stops decreasing (thus validation patience of 1). The training was carried out using both GPUs with an initial learning rate of 0.001. Training lasted for 13hr 40 mins with a validation accuracy of 99.01%. There were a total of 44.4 million learnable parameters. We labelled the trained networks as Stage 1, Stage $2_1$ and Stage $2_2$, (see Table 1).

### 4. Results and Discussion
**4.1** *Classification of Test set*

As stated above, the Test set was neither labeled nor grouped into classes – since the radii were generated with radius-square-law they were naturally sorted in descending order before the images were generated and named sequentially. We used the networks, Stage 1 and $2_2$, to predict the class of each image consecutively. The predicted class-labels were obtained and their individual class-average-radius was calculated from the lower- and upper-class-limits of each class. Then, the class-average-radius was plotted against time, with a time step of 1 s. In estimating the quality of classification, we relied on the fundamental property that each measured radius over the entire evolution – $R(t)$ point – should belong to the same smooth curve. Thus, both the number of outliers and especially the discontinuities of the curve (which form "branches") should be minimal [37]. This suggests the nomenclature that the "*main-branch*" of evaporation refers to the smooth curve that can (ideally) be traced from the initial radius of the droplet to the end of evaporation. Any other curve (usually shorter) that falls above or below this *main-branch* would be referred to as *"upper-branch"* or *"lower-branch"* respectively. Henceforth, we shall adhere to this nomenclature. Figure 6 (a) and (b) represent the networks predictions of the Test set for Stage 1 and $2_2$ respectively. In both plots, only the *main-branch* of evaporation can be observed. Also, it can be inferred from the inset in both Figure 6 (a) and (b) that, multiple images – in the sequence – have been classified into one class by both networks. This is very much to be expected because, as stated earlier, all radii (hence image) generated within 30 nm (for Training data 1) and 10 nm (for Training data 3) interval belonged to one class. This renders multiple images in the sequence to be different by only a fraction of nanometer (e.g. 0.015 nm in Training data 1 and 0.005 nm in Training data 2 in the case of linearly generated radii). In case of images obtained from experiments, this minute difference is not only fundamentally irretrievable from two consecutive images, even with off-line techniques, but is physically meaningless due to such factors as droplet non-sphericity, surface oscillations, finite width of gas-liquid interface and limited knowledge of local values of droplet refractive index. There are, however, two improvements worth revealing in the classification of the Stage $2_2$ network. Firstly, a closer look at the inset in Figure 6 (b) reveals that a smaller number (min = 3 and max = 6) of images have been classified into one class. This improvement is obviously due to the reduction of the class-interval from 30 to 10 nm and generating 1000 images within a smaller radii range thereby increasing the accuracy of the network. Secondly, some of the previously misclassified images have been correctly classified (compare Figure 6 (a) and (b)). This may also be attributed to the increase in accuracy of the network. Both improvements are very promising for our technique in classifying experimental data which may be much more similar consecutively for slowly varying droplets.

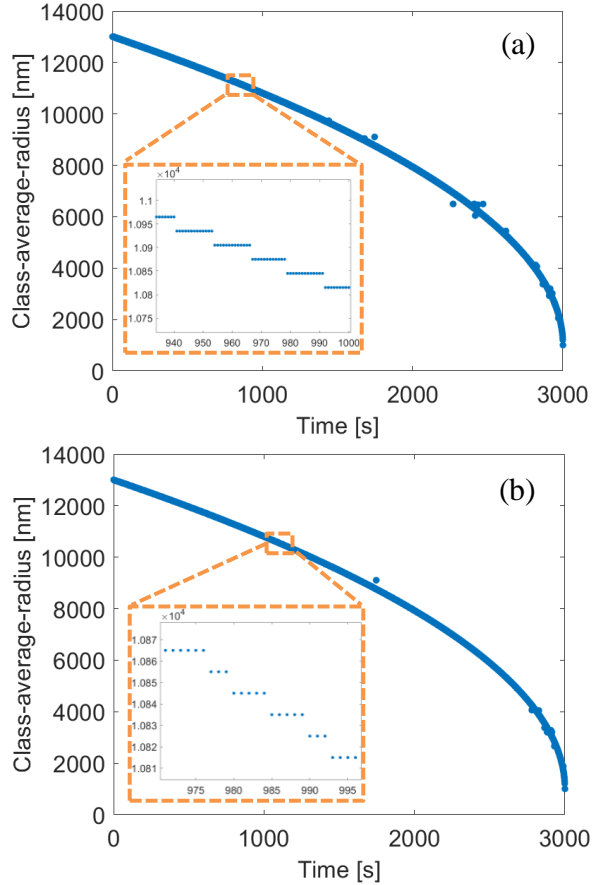

Figure 6: Classification of Test set images corresponding to the time-varying radius of the sphere generated using radius-square-law (see equation (1)). (a) Result of classification using Stage 1 network with 30 nm class interval, (b) result of classification from Stage $2_2$ network with 10 nm class interval. Insets indicate the number of images classified to consecutive classes (accuracy).

**4.2** *Classification of images from microdroplets of pure liquid*

For the pure DEG, we recorded experimental images from two evaporating microdroplets labeled Droplet 1 and Droplet 2 respectively. The images from each droplet were classified using the Stage 1 network, followed by the Stage $2_2$ network. As previously, the networks classified the unlabeled images consecutively. Figure 7 shows plots of predicted class-average-radii (in nm) versus time (in s). The class-average-radius was calculated as described above for the Test set, with the time step (0.0133 s) derived from the camera's frame rate. The droplet label and the network used for classification are displayed on each plot for clarity and comparison. In all plots, the *main-branch* of evaporation – marked with an orange arrow – is clearly visible, demonstrating the networks' ability to track droplet evolution. However, it is also apparent that the Stage 1 network performed poorly as the droplet approached the end of its evolution. This may be attributed to the too large class interval in comparison to the droplet radius together with comparatively worse signal-to-noise ratio of the scattering patterns. Even for pure liquids at hypothetical 0.999 purity, when the droplet's radius changes from 10 to 1 μm, the volume reduces by about $10^3$ times, causing a 0.001 impurity to become significant. Our liquids were of mere 0.99 purity. This also means that, the unknown but non-negligible change of the refractive index, which wasn't accounted for in the training set and, together with other factors, may cause misclassifications at the end of evolution. Such effects may also be expected at the beginning of droplet evolution, if the droplet is contaminated with moisture.

On the contrary, the Stage $2_2$ network with a smaller class interval performed very well even at the end of evolution where dozens of images were classified to one class by the Stage 1 network (compare plots for Stage 1 with Stage $2_2$ network in Figure 7). The plot of Droplet 1, classified with Stage $2_2$ network reveals, at the tail-end of evolution, a trend which signifies the presence of non- or low-volatile impurities. The change of the refractive index must have been small enough not to disrupt the classification. On the other hand, appearance of a *lower-branch* of evaporation at the beginning of evolution (indicated by the green arrow in the bottom-left panel in Figure 7) seems to reveal the presence of contaminating water, which alters the droplet's refractive index – as discussed above. However, it must be pointed out that the thermodynamic model predicts a much faster evaporation rate in that case, as water has higher volatility than DEG. Ultimately, we would like this difference to be clearly manifested in the evaporation curves, as shown in our earlier publication [37].

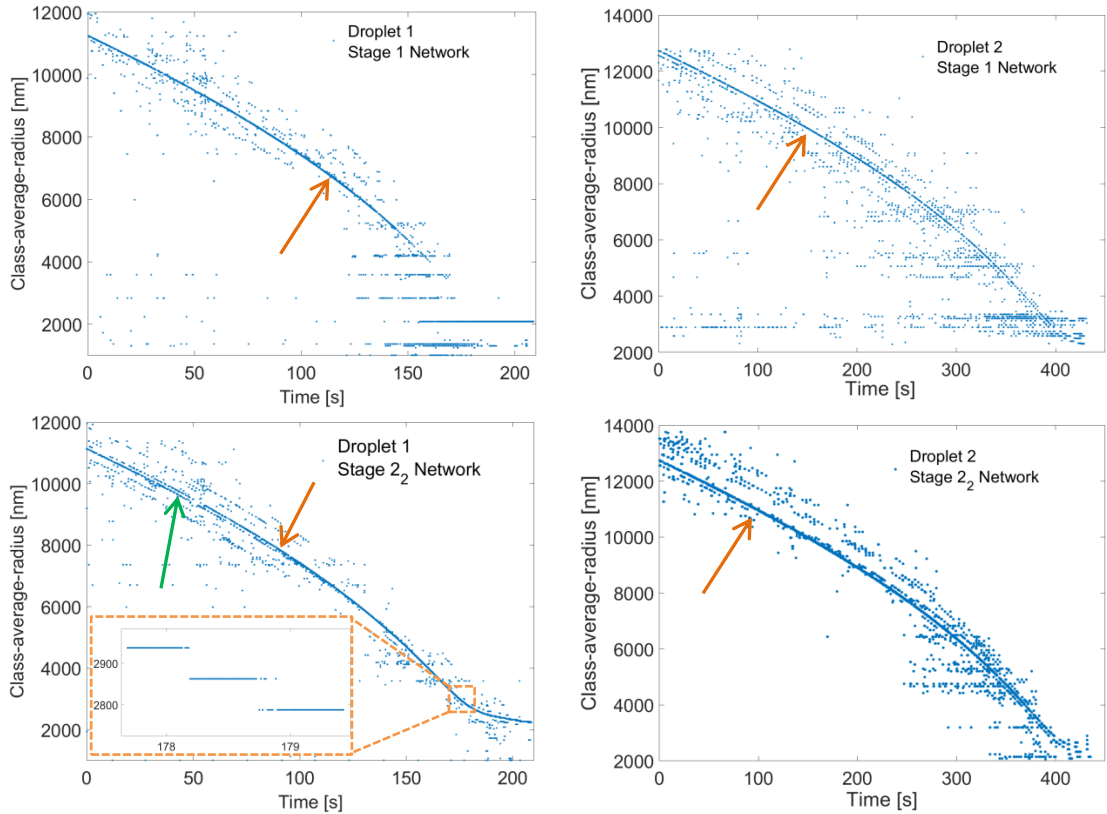

Figure 7: Classification of experimental light scattering patterns recorded from two evaporating microdroplets of pure DEG (Droplet 1 and Droplet 2), yielding droplet radius evolution $R(t)$. The Stage 1 network was trained to cover a wide radii range (1 to 30 μm), divided into 967 classes with a class interval of 30 nm. Stage $2_2$ network covered half of the radii range (1 to 15 μm), divided into 1401 classes with a class interval of 10 nm. In both cases, theoretical data was generated and used for network training. Orange arrows indicate *main-branch(es)* of evaporation – a sequence of $R(t)$ points lying densely on a smooth curve. The *lower-branch* indicated with the green arrow can be attributed to the change of the droplet refractive index associated with the presence of contaminating water. The evaporation was carried out at 21±0.2°C

Regarding the issue of multiple images in a sequence being classified into the same class, we compared the number of such images for the Stage 1 and Stage $2_2$ networks. For both networks, the number of images classified into one class is significantly higher than the corresponding numbers for the Test set described in Section 4.1 (compare insets in Figures 6 and 7). This increase may be attributed to the effect discussed earlier, as well as the high frame rate of the camera relative to the droplet evaporation rate, which causes multiple images in the sequence to appear very similar. To investigate the similarity of these experimental images, the 1st and 20th frames (which are separated by a time interval of 0.266 s) from Droplet 2 were extracted and their mean scattering patterns were obtained and plotted together in Figure 8. As can be seen, only minute variations in the heights of some peaks are visible.

### 4.3 *Classification of images from microdroplets of binary liquids*

For the binary liquids mixture (DPG + $H_2O$), new set of theoretical data was generated using the refractive index of the less volatile component (DPG). We used a class interval of 30 nm and generated 1000 images in each class and retrained the network from scratch. The result of classification of experimental images recorded from a droplet is presented in Figure 9. Three groups of branches with branch discontinuities (indicated by orange circles) are visible. As mentioned in Section 4.2, such additional branches may result from changes in the refractive index (due to changes in droplet composition), which are not accounted for in the theory-generated dataset used for network training. The *upper-branch*, above the discontinuity at ~50 s, can be attributed to the presence of the more volatile liquid ($H_2O$) – with lower refractive index – in the mixture. The *lower-branch*, below the discontinuity ~130 s, can be attributed to the influence of impurities from both liquids that manifest towards the end of evolution.

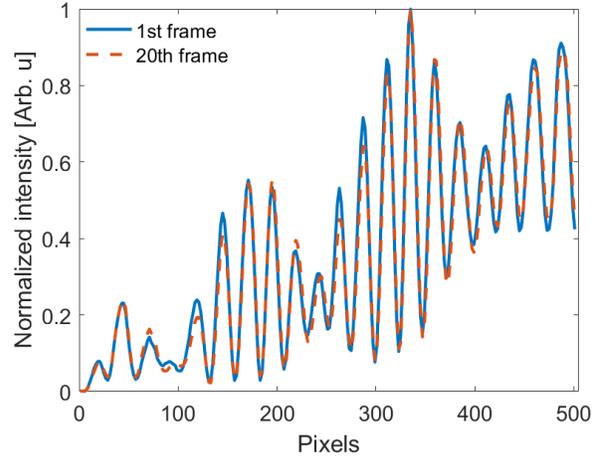

Figure 8: A plot of mean light scattering patterns of the 1st and 20th frames of experimental images recorded from Droplet 2, indicating similarity of patterns.

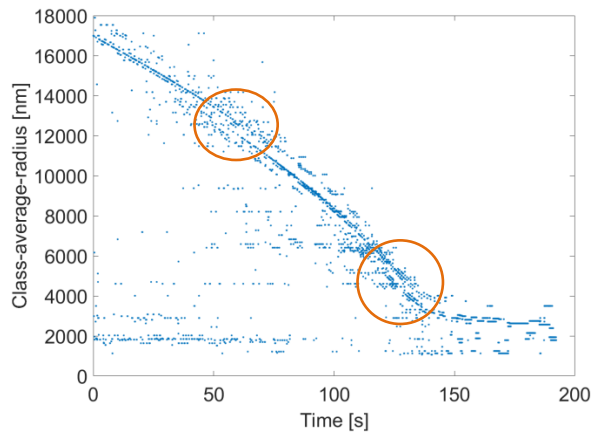

Figure 9: Classification of experimental light scattering patterns recorded from a microdroplet of DPG + $H_2O$ mixture. The network was retrained with a class interval of 30 nm and refractive index of 1.4355 before classification. The evaporation was carried out at 25±0.2°C

As we have outlined in our previous publication, the uncertainty of both droplet position and refractive index are the main factors in droplet radius measurements even with an off-line method [37]. However, it has been noticed that these parameters may vary slightly without destroying the smooth radius evolution curve yielded by this off-line method for pure liquids. Droplets of liquid mixtures and suspensions on the other hand are known to be affected by the variation of the refractive index due to the change of relative content of components and by (varying) inhomogeneity (compare e.g. [9]). In the case of the off-line method we used, this can be overcome by iteratively running the method, augmented with the refractive index change model, as we practiced, for example, with DEG-glycerol mixtures (see [41]). However, when applying a CNN to this task, it remains a training issue – both conceptual and in terms of resources.

*4.4 Classification of images from microdroplets of suspensions*

The experimental images from droplets of the suspension were recorded and processed in the same way as those from pure droplets, as described above. We used the Stage 1 network (trained on pure DEG) to classify the images recorded from the droplets DEG + PS. Meanwhile, the network trained for the DPG + $H_2O$ mixture was used to classify the images of droplets of DPG + $H_2O$ + $SiO_2$ suspension without retraining. In both cases, we assumed that the concentration of PS and $SiO_2$ in the mixtures would have a negligible effect on the refractive index of the suspension. Of course, the assumption was reasonably justified only at the earlier stages of evolution, while the later the worse. The results of classification of images recorded from a droplet of DEG + PS suspension is presented in Figure 10 (a) while the result of DPG + $H_2O$ + $SiO_2$ suspension is presented in Figure 10 (b). As usual, in both figures we present the plot of class-average-radii versus time.

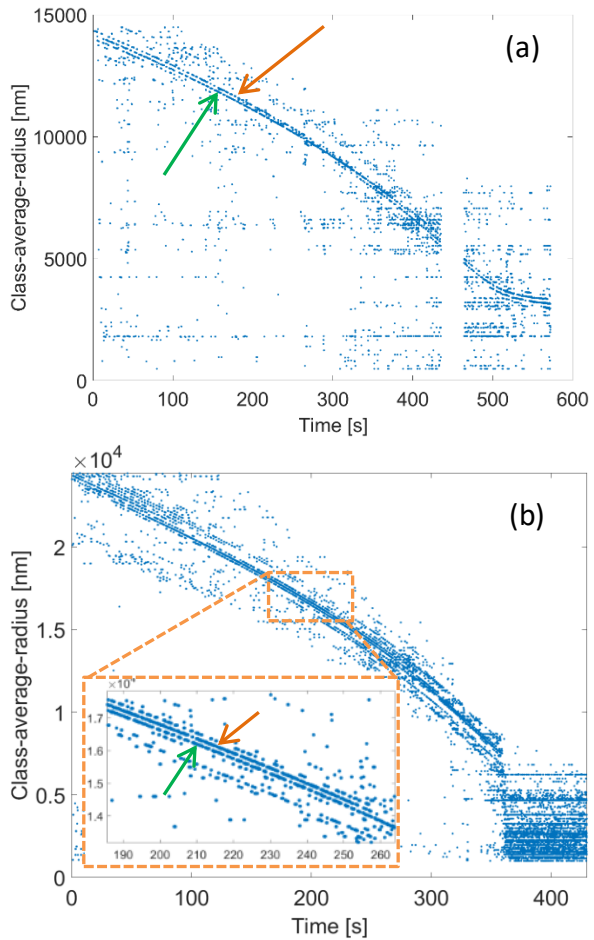

Figure 10: Classification of experimental light scattering patterns recorded from a microdroplet of suspension of polystyrene in diethylene glycol (a) and silica in dipropylene glycol-water mixture (b). The gap in (a) is due to the classification of two consecutive movies from the same droplet. The evaporation was performed at 25±0.2°C. Orange and green arrows indicate *branches* of evaporation.

The influence of ill-chosen refractive index is evident in the higher spread of points and the appearances of multiple branches of evaporation (see orange and green arrows on Figure 10 (a) and (b)). The change in refractive index is associated with: (i) a significant amount of water remnant from original PS and $SiO_2$ suspensions, at the beginning and (ii) increasing number concentration of PS and $SiO_2$ microspheres as the liquids evaporate. For these binary and ternary mixtures, there were at least two *branches* of evaporation, which is in agreement with our earlier experimental findings obtained with off-line methods for mixtures with higher refractive index contrast between components and/or high content of dispersed phase [9–11].

*4.5 Grad-CAM analysis*

Finally, to visualize the CNN's decision-making process, we used one of the state-of-the-art techniques, Gradient-weighted Class Activation Mapping (Grad-CAM, see details in [42]) to generate visual maps (heatmaps) of activated regions in the images for specific target classes. We used this technique to investigate the similarities (or otherwise) of activated regions: (i) within one class (ii) between two consecutive classes (especially at the class boundaries) and (iii) between generated and experimental images (which have been classified into one class). Here, our aim is to unravel the network's discriminative ability between two classes and the recognition ability between generated and experimental images. Henceforth, we will refer to the generated images used for training as theoretical images.

To achieve this, we first selected six predicted-class-labels of experimental images from Droplet 2 (see Table 2) after classifying them using Stage $2_2$ network (see Figure 7). Four out of the six classes were consecutive class-labels (see column 1 in Table 2). The images classified to these classes (see column 2 in Table 2) fell on the *main-branch* of evaporation curve. The images in the other two classes (see columns 4 and 3 respectively in Table 2) have been misclassified (outlier-classes). Hence, these classes constituted our target classes. We used the maps obtained from the four classes (labels: Class 576 – 579) to compare between: (i) consecutive classes of theoretical images (see section 4.5.1) and (ii) theoretical and experimental images (see section 4.5.2). Lastly, the maps obtained from the other two outlier-classes (Class 550 and 582) were used to investigate the misclassification (see section 4.5.3).

### 4.5.1 *Grad-CAM maps of theoretical images*

| Images from main branch of evaporation | | Misclassifications | |
|---|---|---|---|
| Class number | Frame number | Class number | Frame number |
| 576 | 21 – 23 | 550 | 1 –11 |
|  | 27 – 31 |  | 13 – 15 |
|  | 34 – 38 |  |  |
|  | 43 & 44 |  |  |
|  | 50 & 51 |  |  |
| 577 | 79 – 82 | 582 | 12 |
|  | 85 – 91 |  | 16 – 20 |
|  | 93 – 97 |  | 24 – 26 |
|  | 100 – 105 |  | 32 – 33 |
|  | 107 – 220 |  | 39 – 42 |
| 578 |  |  |  |
| 579 | 371 – 376 |  |  |
|  | 379 – 382 |  |  |
|  | 387 & 388 |  |  |

Table 2: Movie frame numbers from the experiment on pure DEG – Droplet 2 – that were classified (using Stage $2_2$ network) into 6 target classes: 4 from the *main-branch* of evaporation and 2 outlier-classes. The Grad-CAM region-activation maps for these images were generated to elucidate the details of classification mechanism.

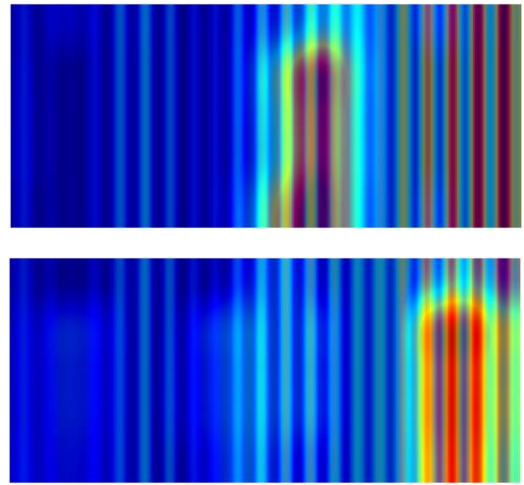

Figure 11: Two sample Grad-CAM region-activation maps superimposed on their respective images (Mie scattering patterns) generated from class 576 showing different regions of higher activation.

For the first task, we reclassify the entire (1000) theoretical images (which were used for training) in each of the target classes to ascertain that they have been classified to their respective classes. The Stage $2_2$ network was used. Then we generated their Grad-CAM region-activation maps (henceforth, called maps). In Figure 11, we present maps of two theoretical image samples taken from Class 576, superimposed on their respective images (we consistently present the maps in this way). For each generated map, there is at least one unique sensitive region that the network used to classify the image into that class. Although these two images belong to the same class, different regions have been activated. This is expected because the images correspond to different radii, which have been grouped into a single class.

The maps of the entire 1000 images from Class 576 and 577 have been tiled together and presented in the Appendix (see Figure A1 and A2). A comparison of the images within one class and between classes show that there are several groups of images within each class with similar maps – a somewhat different region in each group has been strongly activated by the network. It is worth noticing that each such group of images corresponds to an extremely small range of radii – very few nanometers (see for example red dashed rectangles in Figure A1 and A2 in Appendix). This indicates a high sensitivity of the CNN, which could be harnessed to obtain even higher accuracy.

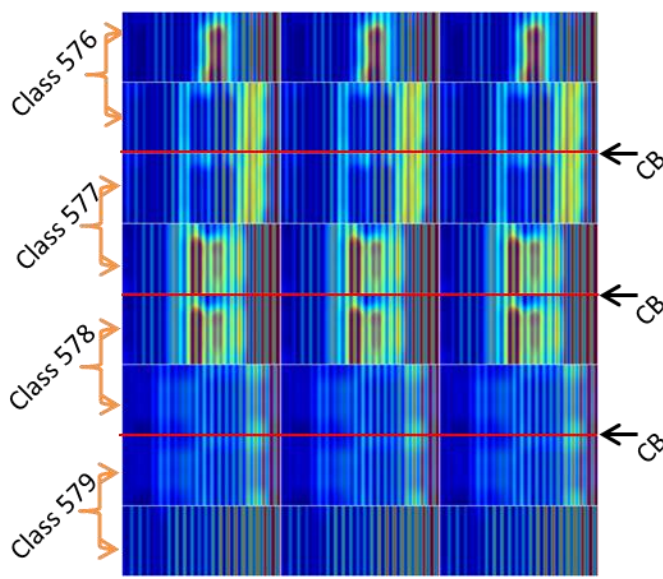

Figure 12: Grad-CAM maps of theoretical images at the class boundaries of the target classes, superimposed on their respective images. The images represent the first-three and the last-three consecutive images of the four target classes, revealing the similarities between images at the class boundaries (CB).

Besides that, we compared the maps of the images at the class boundaries of the four target classes. Figure 12 represents the maps of the first-three and the last-three consecutive images of our target classes. The class boundaries (CB) are separated by red lines and indicated with arrows on the right. The similarities between maps of adjacent classes at their class boundaries can be easily seen (compare maps at the immediate top and bottom of each CB in Figure 12). These results are well to be expected in our technique because consecutive images, artificially separated by the class boundary, are still expected to be very similar. The radius step across the class boundary is just a fraction of nanometer. In such cases, an image may easily be classified to either one of the classes.

#### 4.5.2 *Well-classified experimental images*
For the second task, we produced the maps of the experimental images from Droplet 2. These were the images classified to the four target classes (except Class 578 where no experimental image was classified to) – see Table 2. Here, "*well-classified*" refers to all experimental images which have been classified to the *main-branch* of evaporation curve. We compared these maps with the maps of the theoretical images in their corresponding classes. For each class (labels: Class 576, 577 and 579), we found a group of theoretical images whose maps completely matched the maps of the experimental images. The top panel in Figure 13 represents the 8 theoretical images which were found to match the experimental images (presented in the middle panel) classified to Class 576. In the bottom panel of Figure 13 their mean scattering patterns have been plotted together for easy comparison. The comparison of respective maps for Class 577 and 579 have been presented in Figure A3 and A4 in the Appendix. Again, samples of images from theoretical and experimental images have been presented alongside their mean scattering patterns for comparison in these figures.

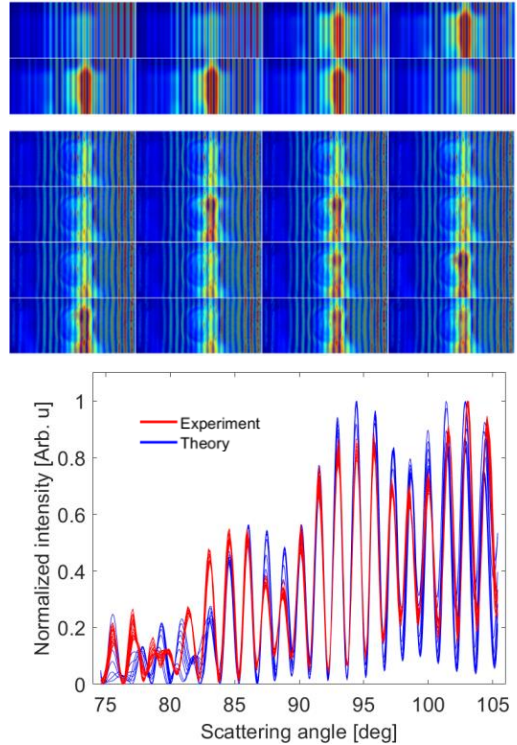

Figure 13: Grad-CAM maps of theory (top panel) and experiment (middle panel) for comparing the activated region for decision making by the network in Class 576. The bottom panel shows a plot of their mean scattering patterns.

### 4.5.3 *Ill-classified experimental images*

Finally, the maps of the images in the two outlier-classes have also been compared with the theoretical images in their respective classes. "*Ill-classified*" refers to all experimental images which have been classified to the outlier-classes. Unlike the "*well-classified*" experimental images discussed above, no group of experimental images could be found to match the theoretical images when their maps were compared. In Figure 14, the only seemingly similar maps have been presented – the top-left panel represents maps of theoretical images while the bottom-left, the maps of experimental images from Class 550. When their mean scattering patterns were obtained and plotted together, they show some level of similarity at the centermost part (see the right panel in Figure 14). However, there exist mismatches of several peaks towards the left and right ends of the patterns which reveal that they have been misclassified into that class.

It should be noted that differences between theoretical and experimental images may be also quite naturally encountered at morphology dependent resonances. The positions and profiles of MDRs are very susceptible to droplet non-sphericity and inhomogeneity [43–45], which is usually the case in experimental setting. Only for perfectly spherical, homogeneous droplet with a constant linear refractive index, these resonances are simply associated with specific size parameters, which can be found using Mie theory [43,44]. Thus, fitting Mie theory predictions to experimental (real-world) data at an MDR may lead to significant discrepancies (compare e.g. [9]).

Very similar effects are expected to appear in classification with CNN. MDRs manifest only on a very few (if any) images used for training (as indicated with orange dashed rectangle in Figure A2 in the Appendix) – "long tail" effect – and thus won't be learned properly.

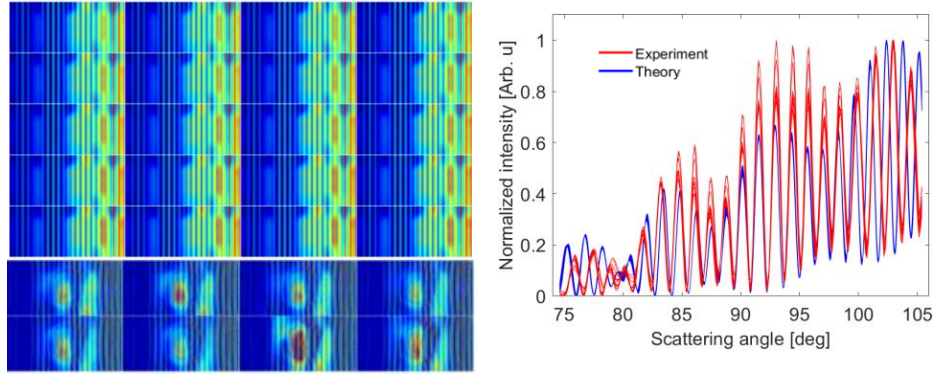

Figure 14: Grad-CAM maps of the theory (top-left panel) and experiment (bottom-left panel) for comparing the activated region for decision making by the network in Class 550. The right panel shows a plot of their mean scattering patterns. The mismatches in the positions of the peaks at the left and right ends of the scattering pattern reveal the misclassification.

Training a CNN to recognize resonances (which, as mentioned earlier, appear significantly different from neighboring images) would actually require creating dedicated classes, which seems not feasible for the real-world data.

We have also investigated the influence of the Newton ring, seen at the center of the experimental images, on the classification accuracy. That is, by summing a large number of the experimental images we obtained a mean image, in which only the static ring is visible, while the moving fringes are averaged-out. This mean image was then normalized to the same range as the experimental images. Firstly, we subtracted this mean image from each experimental image before classification, but there was no clear evidence of improvement in the results. Secondly, instead of multiplying the theoretical images by the random noise (as explained above), we multiplied them by that mean image and trained the network on such dataset. However, this approach did not show improvement in the results either. Nevertheless, we want to try a replacement for the AR-coated (aspherized) lens with an uncoated one to eliminate the ring completely.

## 5. Conclusion

We propose a convolutional neural network technique for online tracking of the temporal evolution of the radius of levitating evaporating pure or composite microdroplets. While we consider our current technique as a first approach, the network is able to process an image (221 x 501 x 1) at ~19 ms (corresponding to 52 fps). We consider the obtained results for the pure droplets as satisfactory, while reliable measurement of a composite droplet radius requires overcoming the problem of concurrent refractive index change/evolution. An engineering solution (especially for online applications) could be to employ a Kalman filter at this stage. We expect that some progress can be gained by extending our two-tier approach – after coarse grained estimation, CNNs trained to more specific ranges of radii and refractive indices would be employed. A better optical system should also help. We envisage that the precision and speed of our network and tech-nique can be further enhanced by e.g. classification of light scattering patterns simultaneously recorded from droplets illuminated with two laser beams of different polarizations and/or color and perhaps simultaneously reducing the vertical dimension of the input images.

## 6. Acknowledgements

This research was funded in whole or in part by National Science Centre, Poland, grant 2021/41/B/ST3/00069. For the purpose of Open Access, the author has applied a CC-BY public copyright license to any Author Accepted Manuscript (AAM) version arising from this submission.

## 7. Appendix

Figure A1 and A2 represent the tiled images of the target Class 576 and 577 respectively, described in section 4.5.1, with Grad-CAM maps superimposed. These are presented here so that the reader may compare the groups of images with similar activation in one class and between the two classes. The group of theoretical images whose maps matched the maps of the experimental images has been indicated with a red-dashed rectangle in the tiled maps. A narrow (~0.03 nm-wide) MDR is indicated with the orange dashed rectangle. Samples of the indicated maps with their plotted mean scattering patterns have been presented in Figure A3 and A4

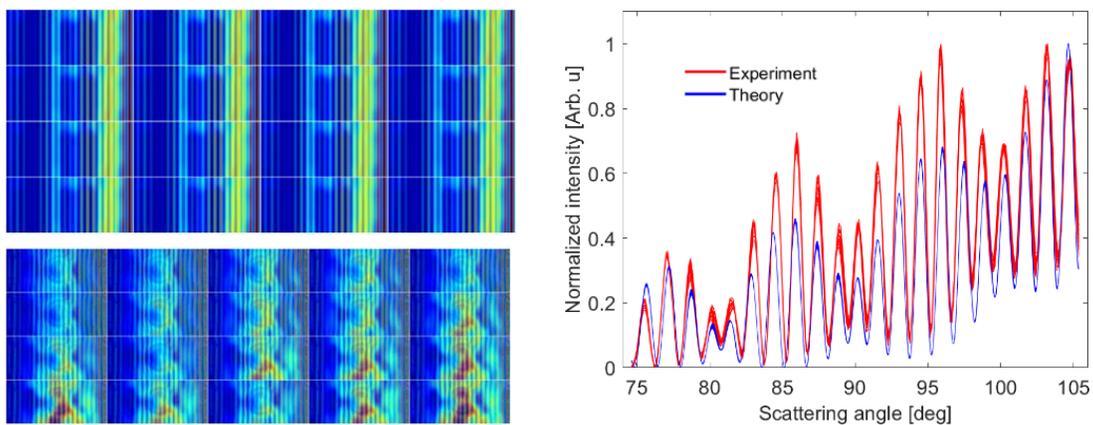

Figure A3: Grad-CAM maps of theory (top-left panel) and experiment (bottom-left panel) for comparing the activated region for decision making by the network in Class 577. The right panel shows a plot of their mean scattering patterns.

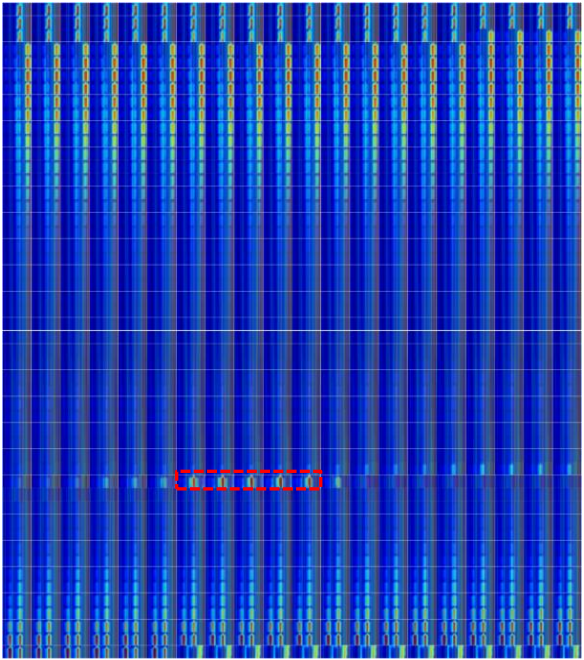 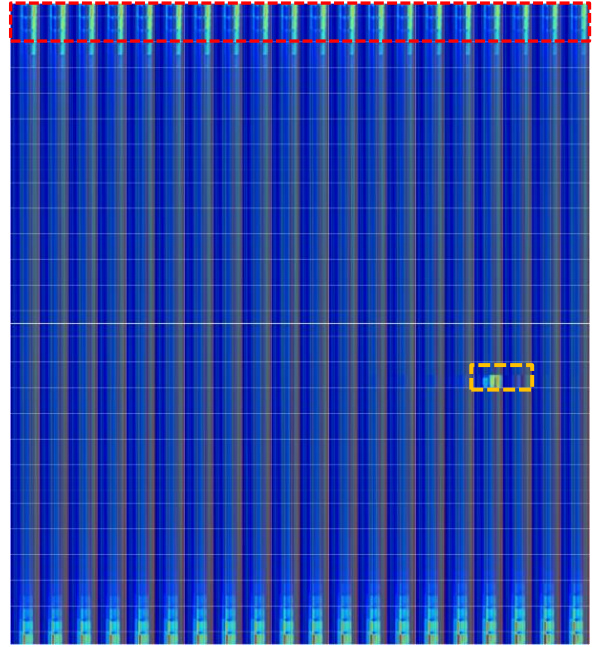

Figure A1: A tiled image of Grad-CAM maps of the 1000 consecutive theoretical images in Class 576. The maps indicated with the red dashed rectangle matched the maps of the experimental images classified to this class.

Figure A2: A tiled image of Grad-CAM maps of the 1000 consecutive theoretical images in Class 577. The maps indicated with the red dashed rectangle matched the maps of the experimental images classified to this class. The orange dashed rectangle indicates a narrow (~0.03 nm-wide) MDR observed in this radius interval.

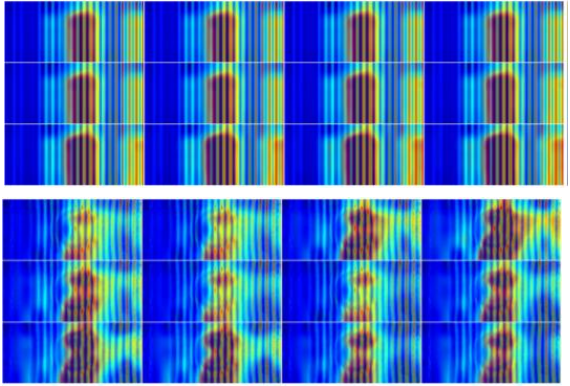 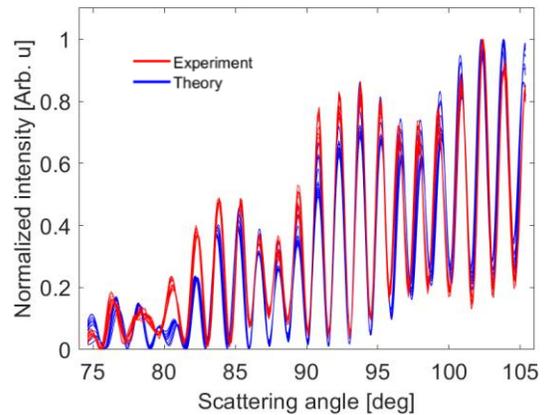

Figure A4: Grad-CAM maps of theory (left-top panel) and experiment (left-bottom panel) for comparing the activated region for decision making by the network in Class 579. The right panel shows a plot of their mean scattering patterns.